# Modeling interfacial phonon transport with normal mode dynamics.

Andrew Rohskopf[a]

[a]Massachusetts Institute of Technology, Department of Mechanical Engineering, Cambridge, MA 02139, USA

**Abstract.**

Traditional theories of interfacial heat transfer by atomic vibrations, also known as phonons, do not explain how vibrational mode interactions contribute to interface conductance. Traditional methods also use the concept of phonons as particles or waves which propagate and transmit energy through interfaces; such methods are therefore inapplicable to realistic non-crystalline systems where phonons are not propagating. Here we introduce a more general formalism of interfacial phonon transport, rigorously derived from interatomic interactions projected onto the normal modes of the system, showing for the first time how interactions between vibrational modes contribute to thermal interface conductance. This new physical picture is based on the concept of forces and energy exchange between modes, regardless of their propagating or non-propagating character, thus providing a general formalism to describe phonon transport in all solids regardless of disorder.

**Introduction.**

Interfacial heat transfer by atomic vibrations is ubiquitous in a variety of phenomena, yet its mechanisms remain elusive. Traditional theories are based on the concept of phonons, which are traditionally thought of as combinations of normal modes in an ideal crystal that form wave-packets, colliding with interfaces like particles and then transmitting their energy. Such theories are rife with assumptions based on intuition and analogy; the commonly used acoustic mismatch model for interface conductance assumes that no scattering occurs at interfaces[1], while the diffuse-mismatch model of interface conductance assumes that all phonons diffusely scatter at interfaces[1], but the actual character of phonon scattering at interfaces is unknown. Methods such as the atomistic Green's function[2] and molecular dynamics (MD)[3,4] only calculate mode contributions to conductance or transmission, but do not elucidate the scattering pathways or mechanisms involved in interfacial heat transfer. The point here is that the physical mechanisms responsible for interfacial heat transfer are elusive, and no formalisms or methods exist to explain interfacial heat transfer in terms of scattering pathways; we seek to remedy this situation by introducing a formalism and method to directly calculate the pathways of vibrational energy transport across interfaces. Our formalism is general in the sense that it can be applied to any solid-solid interface regardless of structural disorder.

We start by introducing a novel atomistic expression for heat/power transfer between two regions of space, such as two sides of an interface, in normal mode coordinates. Using molecular dynamics (MD) simulations, we show that our power transfer expression is consistent and correct, and that harmonic interactions between the modes dominate the interfacial heat transfer. We therefore use the harmonic part of our power transfer expression to analytically solve the Green-Kubo (GK) formula for interface conductance by using the quasi-harmonic Green-Kubo (QHGK) approximation[5], thus yielding novel expressions for interface conductance in the high-temperature classical limit as well as low-temperature quantum regime.

**Instantaneous power transfer in modal coordinates.**

We begin with the general expression for instantaneous power transfer $\dot{Q}_{AB}(t)$ at time $t$ across an interface[6], or generally between two regions of space, with sides denoted by A and B,

$$\dot{Q}_{AB}(t) = -\sum_{i \in A} \sum_{j \in B} \left\{ \mathbf{v}_i \cdot \left( \frac{-\partial H_j}{\partial \mathbf{r}_i} \right) + \mathbf{v}_j \cdot \left( \frac{\partial H_i}{\partial \mathbf{r}_j} \right) \right\} \tag{1}$$

where the sums run over atoms $i$ and $j$, and $\mathbf{v}_i$, $H_i$, and $\mathbf{r}_i$ is the velocity, Hamiltonian, and position of atom $i$. In solids with atoms vibrating about equilibrium, the potential energy of the Hamiltonian takes the form of a Taylor expansion in atomic displacements, so that the spatial derivatives of the single-atom Hamiltonians are

$$\frac{\partial H_j}{\partial u_i^\alpha} = \frac{1}{2} \Phi_{ij}^{\alpha\beta} u_j^\beta + \frac{1}{3} \sum_{k\gamma} \Psi_{ijk}^{\alpha\beta\gamma} u_j^\beta u_k^\gamma + \cdots \tag{2}$$

where $u_i^\alpha$ is the displace of atom $i$ in the $\alpha$ Cartesian direction, $\Phi_{ij}^{\alpha\beta}$ is the harmonic interatomic force constant (IFC2) between atoms $i$ and $j$ in the $\alpha$ and $\beta$ Cartesian directions, and $\Psi_{ijk}^{\alpha\beta\gamma}$ is the harmonic interatomic force constant (IFC2) between atoms $i$ and $j$ in the $\alpha$ and $\beta$ Cartesian directions. Substituting Equation 2 into Equation 1 gives an expression for $\dot{Q}_{AB}(t)$ in terms of atomic displacements and velocities, which can now be transformed to normal mode coordinates using the transformation

$$u_i^\alpha = \frac{1}{\sqrt{m_i}} \sum_n X_n e_{ni}^\alpha \qquad (3)$$

for atomic displacements and

$$v_i^\alpha = \frac{1}{\sqrt{m_i}} \sum_n \dot{X}_n e_{ni}^\alpha \qquad (4)$$

for velocities, where $M_i$ is the mass, and the sum is over all modes $n$ with mode amplitude $X_n$ or momentum $\dot{X}_n$, and $e_{ni}^\alpha$ is the mode $n$ eigenvector component of atom $i$ in the $\alpha$ Cartesian direction.

Using the normal-mode coordinate transformations in Equations 1 and 2, the power transfer $\dot{Q}_{AB}(t)$ may be written in mode coordinates as

$$\dot{Q}_{AB}(t) = \frac{1}{2} \sum_{nm} K_{nm}^{AB} X_n \dot{X}_m + \frac{1}{3} \sum_{nml} K_{nml}^{AB} X_n X_m \dot{X}_l + \cdots \qquad (5)$$

where the sums are over all modes $n, m, l$ and the constants $K_{nm}^{AB}$ and $K_{nml}^{AB}$ are 2nd and 3rd order spatial mode coupling constants (SMCC2s and SMCC3s), respectively, representing the degree of eigenvector overlap between two regions of space $A$ and $B$. In this sense, the coupling constants have intuitive meaning by realizing that $K_{nm}^{AB} X_n$ is the net harmonic force that mode $n$ atoms in side $A$ exert on mode $m$ atoms in side $B$. The quantity $K_{nm}^{AB} X_n \dot{X}_m$ is therefore the net power transfer to mode $m$ atoms in side $B$ from mode $n$ atoms in side $A$; this is a new physical picture of interfacial phonon transport that describes energy exchange between all types of modes, regardless of their propagating or non-propagating behavior.

The 2nd order (harmonic) spatial coupling constants take the form

$$K_{nm}^{AB} = \sum_{i \in B} \sum_{j \in A} \sum_{\alpha\beta} \frac{\Phi_{ij}^{\alpha\beta}}{\sqrt{M_i M_j}} e_{im}^{\alpha} e_{jn}^{\beta} - \sum_{i \in A} \sum_{j \in B} \sum_{\alpha\beta} \frac{\Phi_{ij}^{\alpha\beta}}{\sqrt{M_i M_j}} e_{in}^{\alpha} e_{jm}^{\beta} \tag{6}$$

for the SMCC2s and

$$K_{nml}^{AB} = \sum_{\substack{i \in B \\ j \in A \\ k}} \sum_{\alpha\beta\gamma} \frac{\Psi_{ijk}^{\alpha\beta\gamma}}{\sqrt{M_i M_j M_k}} e_{in}^{\alpha} e_{jm}^{\beta} e_{kl}^{\gamma} - \sum_{\substack{i \in A \\ j \in B \\ k}} \sum_{\alpha\beta\gamma} \frac{\Psi_{ijk}^{\alpha\beta\gamma}}{\sqrt{M_i M_j M_k}} e_{in}^{\alpha} e_{jm}^{\beta} e_{kl}^{\gamma} \tag{7}$$

for the SMCC3s, where the difference between the left and right sums are which side (A or B) the atoms are summed over. We note that the SMCC2s are skew-symmetric:

$$K_{nm}^{AB} = -K_{mn}^{AB} \tag{8}$$

which will be a useful tool when deriving interface conductance later. There are many other ways to represent the power transfer $\dot{Q}_{AB}(t)$ in mode coordinates, but we found this to be the most compact way that minimizes the number of constants (SMCC2s and SMCC3s) that must be calculated.

To show the validity of Equation 5 we perform MD simulations of a silicon-germanium (Si/Ge) interface, where side A is silicon and side B is germanium and show that the time-integrated mode power transfer of Equation 5 agrees with the total energy change of one side of an interface. The agreement of these quantities, along with the observation that the anharmonic terms of Equation 5 represent only a small contribution to the total energy transfer, is shown in Figure 1.

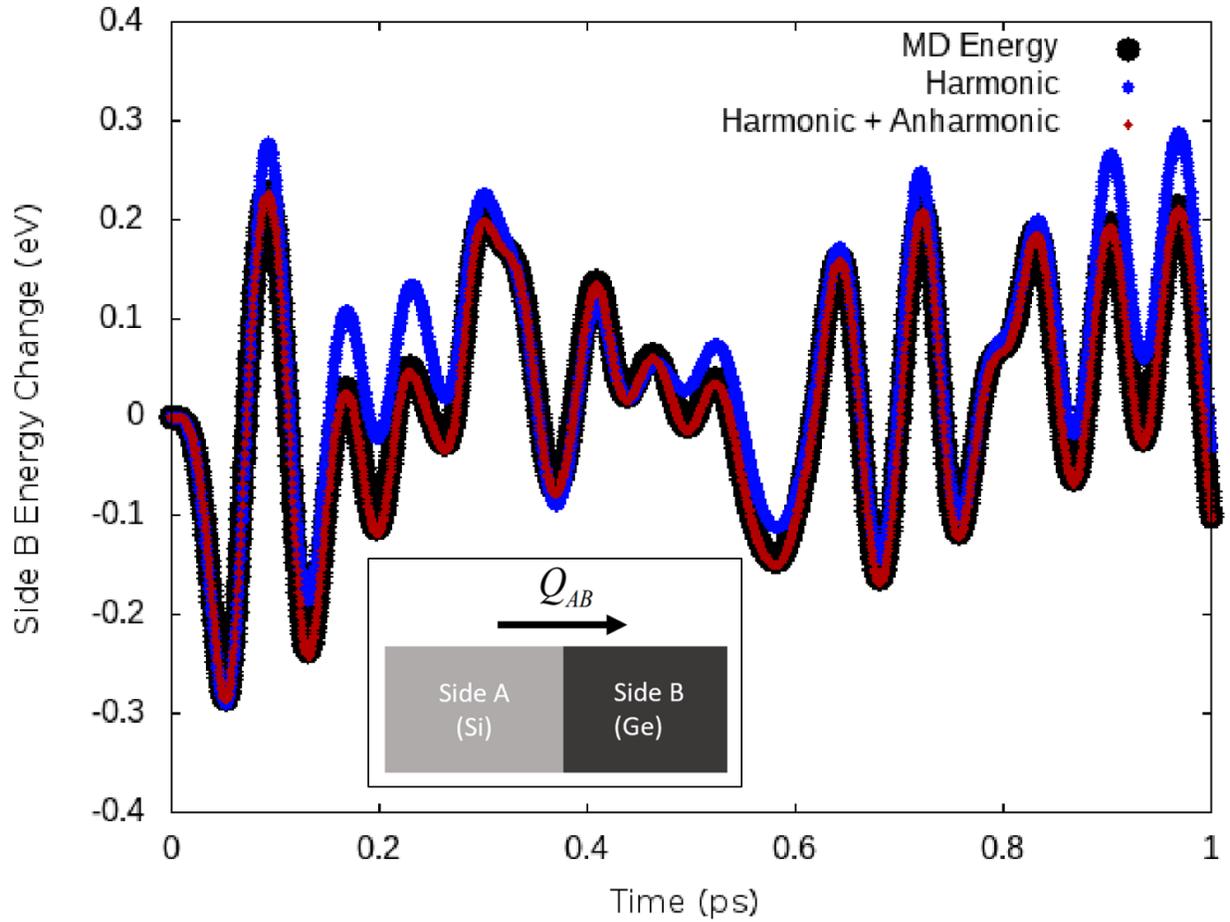

Figure 1. Energy change of germanium (Side B) in a Si/Ge superlattice, due to energy transfer with silicon (Side A), in a MD simulation at 300 K. The picture in the box at the bottom of the plot visually shows the flow of heat being calculated in this simulation. The black line is the energy change calculated as a function of time by the interatomic potential in a MD simulation. The blue line is the time-integrated harmonic part of the mode power transfer expression in Equation 5. The red line is the time-integrated mode power transfer expression of Equation 5, including 3[rd] order anharmonic terms. This shows that the mode power transfer expression of Equation 5, along with the spatial mode coupling constants, are accurate and agree with the energy transfers that occur in a MD simulation.

The method of calculating mode-mode interactions via Equation 5 is dubbed normal mode dynamics (NMD), since we are directly calculating inter-mode forces and energy transfers in MD simulations. The NMD formalism presented here, however, is useful beyond MD simulations; our findings that harmonic forces dominate the interfacial heat transfer (Figure 1) suggest that we may analytically solve for thermal interface conductance using linear response theory.

**A quasi-harmonic Green-Kubo model for interface conductance.**

The QHGK approximation, recently developed for thermal conductivity[5], involves using the harmonic part of the energy current operator in a Green-Kubo (GK) formula to calculate a transport coefficient; we follow this approach, but applied to interface conductance instead of thermal conductivity, by using our novel interfacial heat flow formula in Equation 5. The interface conductance $G$ is calculated from the GK formula[7],

$$G = \frac{1}{Ak_B T^2} \int_0^\infty \langle \dot{Q}_{AB}(t) \dot{Q}_{AB}(0) \rangle dt \qquad (9)$$

where $A$ is the cross-sectional area of the interface, $k_B$ is Boltzmann's constant, and $T$ is the system temperature at equilibrium. To analytically solve Equation 9 with our power transfer expression $\dot{Q}_{AB}(t)$ of Equation 5, we ignore anharmonic terms based on our findings in Figure 1, and we make another coordinate transformation using the complex amplitudes $\alpha_n(t) = \sqrt{\frac{\omega_n}{2}} X_n + \frac{i}{\sqrt{2\omega_n}} \dot{X}_n$, where $\omega_n$ is the frequency of mode $n$. Using complex amplitudes, we may rewrite the mode power transfer $\dot{Q}_{AB}(t)$ as

$$\dot{Q}_{AB}(t) = \frac{i}{2} \sum_{nm} v_{nm} \omega_m \left[ \alpha_n^*(t) + \alpha_n(t) \right] \left[ \alpha_m^*(t) - \alpha_m(t) \right] \qquad (10)$$

where the constants $v_{nm} = \frac{1}{2\sqrt{\omega_n \omega_m}} K_{nm}^{AB}$ are determined by the SMCC2s and introduced for mathematical convenience later. We note that Equation 10 is mathematically identical to the heat flux used by Isaeva et al.[5] except our constants $v_{nm}$ have different meaning, so we may follow their procedure to analytically solve the time-integral of the autocorrelation of Equation 10.

Taking the product of the power transfers evaluated at time $t$ and $t=0$ yields

$$\dot{Q}_{AB}(t)\dot{Q}_{AB}(0) = \frac{-1}{4}\sum_{nm}v_{nm}\omega_m v_{pq}\omega_q\left[\alpha_n^*(t)+\alpha_n(t)\right]\left[\alpha_m^*(t)+\alpha_m(t)\right]\left[\alpha_p^*+\alpha_p\right]\left[\alpha_q^*+\alpha_q\right] \quad (11)$$

which contains a 4$^{th}$ order polynomial in complex amplitudes. When taking the ensemble average of Equation 11, we first expand the 4$^{th}$ order polynomial and take the ensemble average of each term using Wick's theorem[5]: $\langle ABCD\rangle = \langle AB\rangle\langle CD\rangle + \langle AC\rangle\langle DB\rangle + \langle AD\rangle\langle BC\rangle$. This results in more tractable ensemble averages, which are readily evaluated using the Hamiltonian $H=\sum_n \omega_n|\alpha_n|^2$. In doing so, one obtains classical ensemble averages such as

$\langle \alpha_n^*(t)\alpha_m(0)\rangle = \delta_{nm}\frac{k_BT}{\omega_n}e^{i\omega_n t}$ in the harmonic limit. To ensure that the time-integral in the GK formula converges it is necessary to include vibrational linewidths $\gamma_n$ as time-constants:

$\langle \alpha_n^*(t)\alpha_m(0)\rangle = \delta_{nm}\frac{k_BT}{\omega_n}e^{i\omega_n t-\gamma_n t}$. Using these substitutions to evaluate the ensemble average of Equation 11, noting the skew-symmetric property of the SMCC2s in equation 6, and calculating the time-integrals, we obtain the time-integral of the interfacial heat flow autocorrelation function as:

$$\int_0^\infty \langle \dot{Q}_{AB}(t)\dot{Q}_{AB}(0)\rangle dt = \frac{(k_BT)^2}{4}\sum_{nm}\frac{\left(K_{nm}^{AB}\right)^2}{\omega_n\omega_m}\tau_{nm} \quad (12)$$

where the relaxation time matrix $\tau_{nm}$ is given by

$$\tau_{nm} \approx \frac{\gamma_n + \gamma_m}{(\gamma_n + \gamma_m)^2 + (\omega_n - \omega_m)^2} \tag{13}$$

in the limit that linewidths are much smaller than the frequencies[5]. The linewidths may be calculated with Fermi's golden rule[8], or any other means of obtaining the relaxation time which is half the inverse of the linewidth. Finally, substituting the time-integral of Equation 12 into the GK formula gives an expression for interface conductance as

$$G = \frac{k_B}{4A} \sum_{nm} \frac{(K_{nm}^{AB})^2}{\omega_n \omega_m} \tau_{nm} \tag{14}$$

We note that this is a classical derivation, and we can derive a similar expression in the low-temperature quantum regime given by:

$$G = \frac{1}{4A} \sum_{nm} c_{nm} \frac{(K_{nm}^{AB})^2}{\omega_n \omega_m} \tau_{nm} \tag{15}$$

where $c_{nm} = \frac{\hbar \omega_n \omega_m}{T} \frac{n_n - n_m}{\omega_m - \omega_n}$ is a modal heat capacity matrix containing Bose-Einstein distribution functions $n_n$.

These expressions for interface conductance are general in the sense that they can be applied to any system, such as a disordered supercell of atoms. To do so, one must first calculate the interatomic force constants, and then calculate the SMCC2s $K_{nm}^{AB}$ via Equation 6. To calculate the relaxation time matrix, one may use linewidths from Fermi's golden rule, which requires 3$^{rd}$ order mode coupling constants $K_{nml} = \frac{\partial U}{\partial X_n \partial X_m \partial X_l}$, where $U$ is the potential energy. One may easily calculate these 3rd order coupling constants for a supercell by transforming a Taylor expansion potential to normal-mode coordinates to find

$$K_{nml} = \sum_{\substack{ijk \\ \alpha\beta\gamma}} \frac{\Psi_{ijk}^{\alpha\beta\gamma} e_{ni}^{\alpha} e_{mj}^{\beta} e_{lk}^{\gamma}}{\sqrt{m_i m_j m_k}} \qquad (16)$$

which gives a convenient way to calculate the linewidths in Fermi's golden rule, provided one has access to the 3rd order force constants. If one is interested in energy transfer between individual modes regardless of an interface, realize that $-K_{nml} X_m X_l \dot{X}_n$ is the power transferred to mode $n$ from two other modes $m$ and $l$, since $-K_{nml} X_m X_l$ is the force on mode $n$ from two other modes $m$ and $l$. In this regard, the content of this paper encompasses a formalism and method to generally calculate interactions between modes, and mode-interaction contributions to interface conductance in a large supercell of atoms, regardless of structural disorder.

**Conclusion.**

We presented, for the first time, an expression for interfacial power transfer (heat flow rate) in terms of interactions between normal modes, Equation 5. This is a general expression that makes no assumptions other than the interatomic forces in the solid-solid interface are well-approximated by interatomic force constants, which is true for most solids. This is a useful expression because it directly shows the contributions of elastic (harmonic) and inelastic (anharmonic) scattering pathways for heat transfer across an interface. The mode coupling constants are likened to inter-mode force constants, showing how modes can put effective forces on each other to transfer energy, thus providing a new physical picture of phonon transport. Using our NMD formalism with MD simulations, we found that harmonic interactions dominate the heat transfer across Si/Ge interfaces at room temperature, and future studies will determine if this is a general occurrence in many interfaces of technological interest.

Due to our finding that harmonic interactions dominate interfacial heat flow, we note the applicability of the QHGK approximation recently introduced for calculating thermal conductivity[5]. Using our novel Equation 5 for interfacial heat transfer in normal mode coordinates, we apply the QHGK approximation and derive expressions for thermal interface conductance, Equations 14 and 15 in the classical limit and low-temperature quantum regime, respectively. Future studies will test the generality of the QHGK approximation in studying interfacial heat transfer, and possible routes to engineering interface conductance by tuning the harmonic spatial coupling constants $K_{nm}^{AB}$; we note that our expressions for interface conductance suggest that by maximizing $\left(K_{nm}^{AB}\right)^2$, we may maximize conductance.

**References.**

8　　　　　Fabian, J. & Allen, P. B. Anharmonic decay of vibrational states in amorphous silicon. *Physical review letters* **77**, 3839 (1996).